\documentclass[prc,twocolumn,showpacs,amsmath,amssymb,floatfix]{revtex4}
\usepackage{graphicx,color,dcolumn,booktabs}
\usepackage{lscape}
\usepackage{txfonts}
\usepackage{amssymb}
\usepackage{indentfirst}
\usepackage{youngtab,color}

\begin{document}

\title{Study of $qqqc\bar{c}$ five quark system with three kinds of quark-quark hyperfine interaction}
\author{S. G. Yuan$^{1,2,4}$, K. W. Wei$^3$, J. He$^{1,5}$, H. S. Xu$^{1,2}$, B. S. Zou$^{2,3}$}

\affiliation{1. Institute of Modern Physics, Chinese Academy of
Sciences, Lanzhou 730000, China \\2. Theoretical Physics Center for
Science Facilities, Chinese Academy of Sciences, Beijing 100049,
China\\3. Institute of High Energy Physics,
Chinese Academy of Sciences, Beijing 100049, China\\
4. Graduate University of Chinese Academy of Sciences, Beijing
100049, China \\5. Research Center for Hadron and CSR Physics,
Institute of Modern Physics of CAS and Lanzhou University, Lanzhou 730000, China}

\thispagestyle{empty}

\date{\today}

\begin{abstract}

The low-lying energy spectra of five quark systems $uudc\bar{c}$
($I$=$1/2$, $S$=$0$) and $udsc\bar{c}$ ($I$=$0$, $S$=$-1$) are
investigated with three kinds of schematic interactions: the
chromomagnetic interaction, the flavor-spin dependent interaction
and the instanton-induced interaction. In all the three models, the
lowest five quark state ($uudc\bar{c}$ or $udsc\bar{c}$) has an
orbital angular momentum $L=0$ and the spin-parity
$J^{P}=1/2^{-}$; the mass of the lowest $udsc\bar{c}$ state
is heavier than the lowest $uudc\bar{c}$ state.

\end{abstract}

\maketitle

\section{introduction}
\label{sec:intro}

The conventional picture of the proton and the corresponding
excited states are a bound state of three light quarks $uud$ in
constituent quark model (CQM).
Recently, an new measurement about
parity-violating electron scattering (PVES) in JLab affords new
information about the contributions of strange quarks to the charge
and magnetization distributions of the proton, which provides a
direct evidence of the presence of the multiquark components in the
proton~\cite{acha}. The importance of the sea quarks in the proton
is also found in the measurement of the $\bar{d}/\bar{u}$ asymmetry
in the nucleon~\cite{udbar}.

Theoretically, the systematic investigation of baryon mass spectra
and decay properties in CQM shows large deviations of theoretical
values from the experimental data~\cite{capstick}, such as the large
$N\eta$ decay branch ratio of $N^{*}(1535)$ and the strong coupling
of $\Lambda(1405)$ to the $\bar{K}N$. Riska and co-authors suggested
that the mixtures of three-quark components $qqq$ and the multiquark
components $qqqq\bar{q}$ reduce these discrepancies~\cite{zou, an,
li}. In a recent unquenched quark model, by taking into account the
effects of multiquark components via $^3P_0$ pair creation
mechanism, it is also very encouraging to understand the proton spin
problem and flavor asymmetry~\cite{bijker}. The $qqqq\bar{q}$
components could also be in the form of meson-baryon configurations,
such as $N^{*}(1535)$ as a $K\Sigma$ bound state~\cite{Weise} and
$\Lambda(1405)$ as a $\bar{K}N$ bound state~\cite{or}.

In the early 1980s, Brodsky $et\ al.$ proposed that there are
non-negligible intrinsic $uudc\bar{c}$ components $(\sim 1\%)$ in
the proton~\cite{bro}.  Later the study of Shuryak and Zhitnitsky
show a significant charm component in $\eta^{'}$ also~\cite{shur}.
It is natural to expect the high excited baryons contain a large
hidden charm five quark components too.  Recently, some narrow
hidden charm $N^*_{c\bar{c}}$ and $\Lambda^*_{c\bar{c}}$ resonances
were predicted to be dynamically generated in the $PB$ and $VB$
channels with mass above 4 GeV and width smaller than 100
MeV~\cite{wu,wangwl}. These resonances, if observed, definitely
cannot be accommodated into the frame of conventional $qqq$ quark
models. A interesting question is whether these dynamically
generated $N^*_{c\bar{c}}$ and $\Lambda^*_{c\bar{c}}$ resonances can
be distinguished from penta-quark configuration states~\cite{zou,
an, li}. To distinguish the two hadron structure pictures, it should
be worthwhile to explore the mass spectrum of the $qqqc\bar{c}$
consisted of the colored quark cluster $qqqc$ and $\bar{c}$.

The five quark configuration $qqqq\bar{s}$ and $qqqq\bar{c}$ with
exotic quantum numbers have been extensively studied in the chiral
quark model~\cite{stancu2,cbar,cbar2,geno}, colormagnetic interaction model \cite{lipkin,kl} and
instanton-induced interaction model~\cite{semays}. In this work,
we study the mass spectra of the hidden charm systems $uudc\bar{c}$
and $udsc\bar{c}$ with three types of hyperfine interactions,
color-magnetic interaction $(CM)$ based on one-gluon exchange,
chiral interaction $(FS)$ based on meson exchange, and
instanton-induced interaction $(Inst.)$ based on the
non-perturbative QCD vacuum structure.

This paper is organized as follows. In Section \ref{sec:frame}, we
show the wave functions of five quark states and Hamiltonians for
the three types of interactions. In Section \ref{sec:spectra}, the
mass spectra in the positive and negative sectors are presented. The
paper ends with a brief summary.

\section{The wave function and Hamiltonian}
\label{sec:frame}

As dealing with the conventional three quark model we need the wave functions and
Hamiltonian to study the spectrum.
\subsection{Wave functions of five quark systems}

Before going to hidden-charm five quark  $uudc\bar{c}$ and
$udsc\bar{c}$ systems with isospin and strangeness as $(I,
S)=(1/2,0)$ and $(I, S)=(0, -1)$, respectively, we first consider
the four quark subsystem, which can be coupled to an antiquark to
form a hidden-charm five quark system. We use the eigenvalue method
as given in Ref.\cite{chen} to derive the $udsc$ wave functions of
the flavor symmetry $[211]_F$, $[22]_F$, $[4]_F$, $[31]_F$ and
$[1111]_F$ , which correspond to the $SU(4)$ flavor representation
\textbf{15}, \textbf{20}, \textbf{35}, \textbf{45}, and \textbf{1},
respectively. For these flavor multiplets combined with $\bar{c}$,
the following decomposition of the $SU(4)$ representation into
$SU(3)$ representations can be found,
\allowdisplaybreaks
\begin{eqnarray}
\bf
15\times\overline{4}&=&\bf 8^0+1^0+8^0+1^0+{\overline{3}}^1+6^1+15^1
\\\nonumber&+&\bf {\overline{3}}^{1}+{\overline{3}}^2
+{\overline{6}}^2+{\overline{3}}^1+{3}^{-1}\\
\bf 20\times\overline{4}&=&\bf 8^0+{\overline{10}}^0+{\overline{3}}^1+6^1
\\\nonumber&+&\bf 15^1+3^1+15^1
+{\overline{6}}^{-1}+8^0+6^0\\
\bf35\times\overline{4}&=&\bf10^0+35^0+24^1+6^1+15^2+3^2
\\\nonumber&+&\bf {\overline{3}}^4+{15}^{-1}
+10^0+6^1+3^2+8^3+1^3+1^3\\
\bf 45\times\overline{4}&=&\bf 8^0+10^0+27^0+24^1+6^1+15^1
\\\nonumber&+&\bf{\overline{3}}^{1}+6^1+15^2 +3^2+{\overline{6}}^2+3^2+8^3+1^3
\\\nonumber&+&\bf{15}^{-1}+10^0+8^0+6^1+{\overline{3}}^{1}+3^2,\\
\bf 1\times\overline{4}&=&\bf 1^0+3^1
\end{eqnarray}

where the upper indexes denote the charm number. The decomposition
notations in Ref.~\cite{stan1} are adopted. In the current work we
only consider the hidden-charm five quark system, which means charm
number $C$=$0$. The states lying in the octet $\bf 8^0$ and singlet
$\bf 1^0$ of the $qqqc\bar{c}$ states carry the isospin and
strangeness as $(I, S)$=$(1/2, 0)$ and $(I, S)$=$(0, -1)$ for the
octet, and $(I, S)$=$(0, -1)$ for the singlet, respectively, which
are the states we need. The octet can be derived from  $[211]_F$,
$[22]_F$ and $[31]_F$. The singlet can be derived from $[211]_F$ and
$[1111]_F$. Only can $[4]_F$ symmetry form deculplet when combined
with antiquark $\bar{c}$, which does not contain the isospin and
strangeness quantum numbers we want.  The $uudc$ wave function can
be constructed directly by replacement rules mentioned in
Ref.~\cite{glozmanc}. The explicit form of $uudc$ and $udsc$ wave
functions are relegated to Appendix~\ref{wf4q}. The phase convention
is same as in Refs.\cite{an,li}.

The general expression in the flavor-spin coupling scheme for these
five quark wave functions is constructed as
\begin{eqnarray}
\psi^{(i)}(J, J_z)&=&\sum_{a,b,c,d,e,f}\sum_{L_z,S_z,s_z}\nonumber\\
&&C^{[1^4]}_{[X^{(i)}]_f[CFS^{(i)}]_e}
C^{[CFS^{(i)}]_e}_{[C^{(i)}]_d[FS^{(i)}]_c}
C^{[FS^{(i)}]_c}_{[F]_a [S^{(i)}]_b}\nonumber\\
&\cdot&[X^{(i)}]_{f,L_z}[F^{(i)}]_{a,T_z}
[S^{(i)}]_{b,S_z}\psi^C_{[211]_d} \nonumber\\&\cdot&
(S,S_z,L,L_z|\tilde{J},\tilde{J}_z)(\tilde{J},\tilde{J}_z,1/2,s_z|J,J_z)
\nonumber\\&\cdot&\bar\xi_{s_z}\varphi(r_{\bar{c}})\bar{\psi}^C\bar{\varphi}.
\label{wfc}
\end{eqnarray}  
where $\tilde{J}$ is the total angular momentum of four quark and
$S$ the total spin of four quark, $i$ is the number of the $qqqc\bar
c$ configuration in both positive and negative parity sectors, which
will be given explicitly later. $\bar{\psi}^C$, $\bar{\varphi}$ and
$\bar\xi_{s_z}$ represent the color, flavor and spinor  wave
functions of the antiquark, respectively. $\varphi(r_{\bar{c}})$
represents the space wave function for antiquark. The symbols
$C^{[.]}_{[..][...]}$ are $S_4$ Clebsch-Gordan coefficients for the
indicated color-flavor-spin ($[CFS]$), color $\bar{\psi}^C$,
flavor-spin ($[FS]$), flavor ($[F]$), spin ($[S]$), and orbital
($[X]$) wave functions of the $qqqc$ system.

\subsection{Hamiltonians}

To investigate the mass spectrum of the five quark system, the
non-relativistic harmonic oscillator Hamiltonian is introduced as in
the light flavor case \cite{Helminen}:
\begin{eqnarray}
H&=&\sum_{i=1}^5 (m_i+{\vec p_i^{\ 2}\over2m_i}) - {\vec
P_{cm}^{2}\over2M}\nonumber\\
&+& {1\over2}\sum_{i<j}^5(C[r_{i}-r_{j}]^2+V_0)+H_{hyp}.
\label{ham}
\end{eqnarray}
where $m_i$ denotes the constituent masses of quarks $u,d,s,c$ (and
the antiquark $\bar c$), and $\vec P_{cm}$ and $M$ are the total
momentum and total mass $\sum_{i=1}^5m_i$ of the five quark system.
$C$ and $V_0$ are constants. As pointed out by Glozman and Riska
\cite{glozmanc}, one may treat the heavy-light quark mass difference
by including a flavor dependent perturbation term $H_0^{''}$,
\begin{eqnarray}
H^{'}&=&\sum_{i=1}^5 (m_i+{\vec p_i^{\ 2}\over2m}) - {\vec
P_{cm}^{2}\over10m}\nonumber\\
&+& {1\over2}\sum_{i<j}^5(C[r_{i}-r_{j}]^2+V_0) +H_0^{''}+H_{hyp}.
\label{ham}
\end{eqnarray}
with $m$ denoting the $u,d,s$ quark mass. The Hamiltonian may be
rewritten as a sum of 4 separated hamiltonians in Jacobi
coordinates. The perturbation term $H_0^{''}$ has the following form
\begin{eqnarray}
H_0^{''}&=&-\sum_{i=1}^4 {(1-\frac{m}{m_c})}\{\frac{\vec p_i^{\
2}}{2m}- {m_c\vec
P^{2}_{cm}\over5m(3m+2m_c)}\}\delta_{ic}\nonumber\\
&-&(1-\frac{m}{m_{\bar{c}}})\{\frac{\vec p_5^{\
2}}{2m}\}\delta_{5\bar{c}}, \label{ham2}
\end{eqnarray}
where the Kronecker symbol $\delta_{ic}$ means that the
flavor-dependent term is nonzero when the $i^{th}$ quark of four
quarks is charm quark. If the center-of-mass term is dropped, the
matrix element of perturbation term on the harmonic oscillator state
in the negative parity sector ($L=0$) will be
\begin{equation}
\langle H_0^{''}
\rangle_{[4]_X[1111]_{CFS}[211]_C[31]_{FS}}=-\frac{3}{4}\delta,
\label{grme}
\end{equation}
where $\delta=(1-m/m_c)\omega_5$ with the oscillator frequency
$\omega_5=\sqrt{5C/m}$. For other states considered in this work the
matrix elements can be also written as such simple form.

The term $H_{hyp}$ reflects the hyperfine interaction between quarks
in the hadrons. In this work we consider three types of the
hyperfine interactions, {\sl i.e.}, flavor-spin interaction $(FS)$ based on
meson exchange, color-magnetic interaction $(CM)$ based on one-gluon
exchange, and instanton-induced interaction (\textsl{inst.}) based on the
non-perturbative QCD vacuum structure.

The flavor-spin dependent interaction reproduces well the
light-quark baryon spectrum, especially the correct ordering of
positive and negative parity states in all the considered spectrum
\cite{GlRi}. The flavor dependent interaction has been extended to
heavy baryons sector in Ref.~\cite{glozmanc}. Given that $SU(4)$
flavor symmetry is broken mainly through the quark mass differences,
the hyperfine Hamiltonian can be written as the following
form~\cite{glozmanc,belg}
\begin{equation}
H_{FS}=-C_{\chi}\sum_{i,j}^{4}\frac{m^2}{m_{i}m_{j}}\sum_{F=1}^{14}\vec\lambda_i^{F}\cdot
\vec\lambda_j^{F} \vec\sigma_i\cdot\vec\sigma_j, \label{fs}
\end{equation}
where $\sigma_i$ and $\lambda_i^F$ are Pauli spin matrices and
Gell-Mann $SU(4)_F$ flavor matrices, respectively, and $C_{\chi}$ a
constant phenomenologically 20$\sim $30 MeV.  In the chiral quark
model~\cite{GlRi},  only the hyperfine interactions between quarks
are considered while the interactions between the quarks and the
heavy antiquark $\bar{c}$ are neglected.

The chromomagnetic interaction, which have achieved considerable
empirical success in describing the splitting in baryon spectra
\cite{capsgur}, are intensively used in the study of multiquark
configurations \cite{ha,ho,stan1,stancu3}. A commonly used hyperfine
interaction is as the following~\cite{ho},
\begin{equation}
H_{CM}=-\sum_{i,j}C_{i,j}\vec\lambda_i^{c}\cdot \vec\lambda_j^{c}
\vec\sigma_i\cdot\vec\sigma_j, \label{cs}
\end{equation}
where $\sigma_i$ is the Pauli spin matrice, $\lambda_i^c$ is the
Gell-Mann $SU(3)_C$ color matrices, and $C_{i,j}$ the colormagnetic
interaction strength. The quark-antiquark strength factors are fixed
by the hyperfine splittings of the mesons. For an antiquark the
following replacement should be applied \cite{jaffe}:
$\vec{\lambda^{c}}\rightarrow -\vec{\lambda^{c}}^{*}$.

The instanton induced interaction, introduced first by $'$t~
Hooft~\cite{hooft} for [ud]-quarks and then extended to three flavor
case~\cite{shiff} and four flavor case~\cite{metsch}, is also quite
successful in generating the hyperfine structure of the baryon
spectrum. The nonrelativistic limit of the unregularized quark-quark
$'$t~ Hooft interaction has the form~\cite{metsch,migura,shur2,dey},
\begin{eqnarray}
\label{hooft} H_{Inst} &=& - 4 {\cal P}_{S=0}^{\cal D} \otimes
\big[{\cal W}_{nn}\;{\cal P}_{\cal A}^{\cal F}(nn) + {\cal
W}_{ns}\;{\cal P}_{\cal A}^{\cal F}(ns)\nonumber\\&+&{\cal
W}_{nc}\;{\cal P}_{\cal A}^{\cal F}(nc) + {\cal W}_{sc}\;{\cal
P}_{\cal A}^{\cal F}(sc)\big] \otimes {\cal P}_{\bf \bar 3}^{\cal
C}\nonumber\\&-&2 {\cal P}_{S=1}^{\cal D} \otimes \big[{\cal
W}_{nn}\;{\cal P}_{\cal A}^{\cal F}(nn) + {\cal W}_{ns}\;{\cal
P}_{\cal A}^{\cal F}(ns)\nonumber\\&+&{\cal W}_{nc}\;{\cal P}_{\cal
A}^{\cal F}(nc) + {\cal W}_{sc}\;{\cal P}_{\cal A}^{\cal F}(sc)\big]
\otimes {\cal P}_{\bf 6}^{\cal C},
\label{hooft}
\end{eqnarray}
where ${\cal W}_{f_1f_2}$ is the radial matrix element of the
contact interaction between a quark pair with flavors $f_1$ and
$f_2$, ${\cal P}_{\cal A}^{\cal F}(f_1f_2)$ the projector onto
flavor-antisymmetric quark pairs; ${\cal P}_{\bf \bar 3}^{\cal C}$
and ${\cal P}_{\bf 6}^{\cal C}$ the projectors onto color
antitriplet and color sextet pairs, respectively; ${\cal
P}_{S=0}^{\cal D}$ and ${\cal P}_{S=1}^{\cal D}$ the projectors onto
antisymmetric spin-singlet and symmetric spin-triplet states,
respectively. For a three quark system, only two quarks $qq$ in a
spin singlet state with the flavor antisymmetry can interact through
the instanton induced interaction. Here, we phenomenologically
consider the instanton-induced interaction of the $nc$ and $sc$
quark pairs, although some authors~\cite{hooft2, sach} assume that
the heavy flavor decouples when the quark gets heavier than the
$\Lambda_{QCD}$.

\section{Mass spectra of $uudc\bar{c}$ and $udsc\bar{c}$ systems} \label{sec:spectra}

In this section, we present the numerical results for the low-lying
spectra of the five quark systems of $uudc\bar{c}$ and $udsc\bar{c}$
with the hyperfine interaction given by the color-magnetic
interaction , the flavor-spin interaction, and the instanton-induced interaction, respectively. For the
kinetic part and the confinement potential part of the Hamiltonian,
we take the parameters of Refs.~\cite{Helminen,glozmanc}, {\sl
i.e.}, $m_u=m_d=340$ MeV, $m_s=460$ MeV, $m_c=1652$ MeV and
$C=m_u\omega_5^2/5$ with $\omega_5=228$ MeV.

All other parameters for three different hyperfine interactions are
listed in Table~\ref{para3}. For the $FS$ model, the $C_\chi$
parameter is taken from Ref.~\cite{Helminen}. For the $CM$ model, we
take the $C_{i,j}$ parameters of Ref.~\cite{ho}, determined by a fit
to the charmed ground states. For the $Inst.$ model, the
parameters are determined by a fit to the splittings between the
baryon ground states $N(938)$, $\Delta(1232)$, $\Lambda(1116)$,
$\Sigma^0(1193)$, $\Omega(1672)$, $\Lambda_c(2286)$,
$\Sigma_c(2455)$, $\Xi^{0}_c(2471)$, $\Xi^{'0}_c(2578)$ and
$\Xi^{*0}_c(2645)$. The fit yields a ratio of about ${\cal
W}_{ns}/{\cal W}_{nn} \simeq 2/3$, which is the same as in
Ref.~\cite{metsch}. The parameter $V_0$ for each model is adjusted
to reproduce the mass of $N^*(1535)$ as the lowest $J^P=1/2^-$ $N^*$
resonance of penta-quark nature.

\begin{table}[h!]
\caption{The parameters (in the unit of MeV) for three kinds of
hyperfine interactions. \label{para3}}
\renewcommand\tabcolsep{0.18cm}
\renewcommand{\arraystretch}{1.1}
\begin{tabular}{c|ccccccccc}
\hline
$CM$~\cite{ho} & $C_{qq}$      &20&  $C_{qs}$       & 14 &$C_{qc}$& 4& $C_{sc}$&5\\
  & $C_{q \bar c}$&6.6  &  $C_{s \bar c}$ &6.7
&$C_{c\bar c}$ &5.5&$V_0$&-208 \\
\hline  $FS$~\cite{Helminen} &$C_{\chi}$& 21 &$V_0$& -269
\\\hline
$Inst.$& ${\cal W}_{nn}$&315&${\cal W}_{ns}$& 200 & ${\cal W}_{nc}$&70&${\cal W}_{sc}$&52\\
 &$V_0$&-213 &\\\hline
\end{tabular}
\end{table}

With all these Hamiltonian parameters fixed and the wave functions
of five quark system outlined in the Sect.II, the matrix elements of
Hamiltonian for various five-quark states can be calculated.

For the $uudc\bar c$ and $udsc\bar c$ systems, the lowest states are
expected to have all five quark in the spatial ground state of
$[4]_X$ configuration and hence negative parity. For the
construction of color-flavor-spin wave-functions, the convenient
coupling schemes for the $FS$ and $CM$ models are different, {\sl i.e.},
$[1111]_{CFS}[211]_C[f]_{FS}[f]_F[f]_S$ and
$[1111]_{CFS}[f]_F[f]_{CS}[211]_C[f]_S$, respectively. The
flavor-spin configurations for the $uudc\bar c$ and $udsc\bar c$
systems of spacial ground state $[4]_X$ for the $FS$ and $CM$ models are
listed in Table~\ref{config}, where the configurations $|1'>$ and
$|3'>$ are only for the $udsc\bar c$ system. For the $udsc\bar c$
system,  the $[211]^{'}_F$ and $[211]_F$ correspond to the Weyl
Tableaus \young(uc,d,s) and \young(us,d,c), respectively.

\begin{table}[h!]
\caption{The flavor-spin configurations for the $uudc\bar c$ and
$udsc\bar c$ systems of spacial ground state $[4]_X$ for the $FS$ and
$CM$ models. \label{config}}
\renewcommand\tabcolsep{0.18cm}
\renewcommand{\arraystretch}{1.1}
\begin{tabular}{l|l|l}
\hline & $FS$ model & $CM$ model\\ \hline $|1'>$ &
$[31]_{FS}[211]^{'}_F[22
]_S$ & $[211]^{'}_F[31]_{CS}[211]_C[22]_S$\\
$|3'>$ & $[31]_{FS}[211]^{'}_F[31
]_S$ &$[211]^{'}_F[31]_{CS}[211]_C[31]_S$\\
$|1>$ & $[31]_{FS}[211]_F[22]_S$ &$[211]_F[31]_{CS}[211]_C[22]_S$\\
$|2>$ & $[31]_{FS}[31]_F[22
]_S$ &$[31]_F[211]_{CS}[211]_C[22]_S$\\
$|3>$ &$[31]_{FS}[211]_F[31]_S$ &$[211]_F[31]_{CS}[211]_C[31]_S$\\
$|4>$ &$[31]_{FS}[22]_F[31
]_S$ &$[22]_F[22]_{CS}[211]_C[31]_S$\\
$|5>$ & $[31]_{FS}[31]_F[31]_S$ & $[31]_F[211]_{CS}[211]_C[31]_S$\\
$|6>$ & $[31]_{FS}[31]_{F}[4]_S$ & $[31]_F[211]_{CS}[211]_C [4]_S$
\\\hline
\end{tabular}

\end{table}

The corresponding seven $udsc\bar c$ wave functions with spin-parity
$1/2^-$ are $|1', 1/2^- \rangle$, $|1, 1/2^-
\rangle$, $|2, 1/2^- \rangle$, $|3', 1/2^- \rangle$,
$|3, 1/2^- \rangle$, $|4, 1/2^- \rangle$, and $|5,
1/2^- \rangle$. The five wave functions with spin-parity
$3/2^-$ are $|3', 3/2^- \rangle$, $|3, 3/2^-
\rangle$, $|4, 3/2^- \rangle$, $|5, 3/2^- \rangle$
and $|6, 3/2^- \rangle$. The one wave function with
spin-parity $5/2^-$ is $|6, 5/2^- \rangle$. They
form three subspace of $J^P=1/2^-$, $3/2^-$ and $5/2^-$,
respectively.

The energies for these different configurations have been calculated
with three kinds of hyperfine interactions and are listed in
Table~\ref{Tab:DN}.

\begin{table}[h!]
\caption{Energies (in unit of MeV) of the $udsc\bar c$ and $uudc\bar c$ system
of the spacial ground state with three kinds of hyperfine
interactions for different flavor-spin configurations.
\label{Tab:DN}}
\renewcommand{\arraystretch}{1.2}
\renewcommand\tabcolsep{0.15cm}
\begin{tabular}{lc|rr|rr|rr|r}
\hline  &&\multicolumn{2}{c|}{$CM$} &\multicolumn{2}{c|}{$FS$}
    &\multicolumn{2}{c|}{$Inst$}&
\\\hline
 conf.&&$udsc\bar{c}$&$uudc\bar{c}$ & $udsc\bar{c}$ & $uudc\bar{c}$&
$udsc\bar{c}$ & $uudc\bar{c}$&$J^{p}$
\\\hline
$|1'>$&& $4404$   &  $--$       & $4169$ & $--$   & $4211$  & $--$    & ${\frac{1}{2}}^{-}$\\
$|3'>$&& $4325$   &  $--$       & $4169$ & $--$   & $4222$  & $--$    & ${\frac{1}{2}}^{-}$\\
      && $4432$   &  $--$       & $4169$ & $--$   & $4222$  & $--$    & ${\frac{3}{2}}^{-}$\\
$|1>$&& $4480$    &  $4372$     & $4156$ & $4017$ & $4287$  & $4125$  & ${\frac{1}{2}}^{-}$\\
$|3>$&& $4441$    &  $4333$     & $4200$ & $4059$ & $4322$  & $4167$  & ${\frac{1}{2}}^{-}$\\
     && $4538$    &  $4430$     & $4200$ & $4059$ & $4322$  & $4167$  & ${\frac{3}{2}}^{-}$\\
$|2>$&& $4552$    &  $4436$     & $4182$ & $4052$ & $4347$  & $4195$  & ${\frac{1}{2}}^{-}$\\
$|4>$&& $4471$    &  $4368$     & $4229$ & $4096$ & $4360$  & $4202$  & ${\frac{1}{2}}^{-}$\\
     && $4572$    &  $4468$     & $4229$ & $4096$ & $4360$  & $4202$  & ${\frac{3}{2}}^{-}$\\
$|5>$&& $4617$    &  $4508$     & $4258$ & $4133$ & $4386$  & $4237$  & ${\frac{1}{2}}^{-}$\\
     && $4585$    &  $4477$     & $4258$ & $4133$ & $4386$  & $4237$  & ${\frac{3}{2}}^{-}$\\
$|6>$&& $4629$    &  $4526$     & $4362$ & $4236$ & $4461$  & $4322$  & ${\frac{3}{2}}^{-}$\\
     && $4719$    &  $4616$     & $4362$ & $4236$ & $4461$  & $4322$  & ${\frac{5}{2}}^{-}$\\
\hline
\end{tabular}
\end{table}

For subspaces of $J^P=1/2^-$ and $3/2^-$, some non-diagonal matrix
elements of Hamiltonians are not zero and lead to the mixture of the
configurations with the same spin-parity. After considering the
configuration mixing, the eigenvalues of the Hamiltonians of the
five quark $udsc\bar c$ and $uudc\bar c$ systems in the spatial
ground state are listed in Table~\ref{int3n}. The corresponding
mixing coefficients of the states with spin-parity $1/2^-$ for three
different models are listed in Tables~\ref{coec}-\ref{coef}. The
spin symmetry $[4]_S$ is orthogonal to the spin symmetry $[31]_S$
and $[22]_S$. There is no mixing between the configuration
$[31]_{FS}[31]_F[4]_S$ and other 7 configurations.

\begin{table}[h]
\caption{Energies (in unit of MeV) the $udsc\bar c$ and
$uudc\bar c$ systems in the spatial ground state under three kinds
of hyperfine interactions ({\sl i.e.}, with configuration mixing
considered ). \label{int3n}}
\renewcommand\tabcolsep{0.25cm}
\renewcommand{\arraystretch}{1.2}
\begin{tabular}{c|cc|cc|cc}
    \hline  &\multicolumn{2}{c|}{CM} &\multicolumn{2}{c|}{FS}
    &\multicolumn{2}{c}{$Inst.$}\\\hline
$J^P$& $udsc\bar{c}$ & $uudc\bar{c}$ & $udsc\bar{c}$ & $uudc\bar{c}$
& $udsc\bar{c}$& $uudc\bar{c}$ \\\hline
${\frac{1}{2}}^{-}$ & $4273$ & $4267$  &$4084$ & $3933$   & $4209$ & $4114$\\
${\frac{1}{2}}^{-}$ & $4377$ & $4363$  &$4154$ & $4013$   & $4216$ & $4131$\\
${\frac{1}{2}}^{-}$ & $4453$ & $4377$  &$4160$ & $4119$   & $4277$ & $4204$\\
${\frac{1}{2}}^{-}$ & $4469$ & $4471$  &$4171$ & $4136$   & $4295$ & $4207$\\
${\frac{1}{2}}^{-}$ & $4494$ & $4541$  &$4253$ & $4156$   & $4360$ & $4272$\\
${\frac{1}{2}}^{-}$ & $4576$ & $~~~~$  &$4263$ & $~~~~$   & $4362$ & $~~~~$\\
${\frac{1}{2}}^{-}$ & $4649$ & $~~~~$  &$4278$ & $~~~~$   & $4416$ &
$~~~~$  \\\hline
${\frac{3}{2}}^{-}$ & $4431$ & $4389$  &$4184$ & $4013$   & $4216$ & $4131$\\
${\frac{3}{2}}^{-}$ & $4503$ & $4445$  &$4171$ & $4119$   & $4295$ & $4204$\\
${\frac{3}{2}}^{-}$ & $4549$ & $4476$  &$4263$ & $4136$   & $4362$ & $4272$\\
${\frac{3}{2}}^{-}$ & $4577$ & $4526$  &$4278$ & $4236$   & $4416$ & $4322$\\
${\frac{3}{2}}^{-}$ & $4629$ & $~~~~$  &$4362$ & $~~~~$   & $4461$ &
$~~~~$  \\\hline
${\frac{5}{2}}^{-}$ & $4719$ & $4616$  &$4362$ & $4236$   & $4461$ & $4322$\\
\hline
\end{tabular}
\end{table}

\begin{table}[h!]
\caption{The mixing coefficients of the states with spin-parity
$1/2^-$ under the $CM$ interaction including the
$q\bar q$ interaction. \label{coec}}
\renewcommand\tabcolsep{0.18cm}
\renewcommand{\arraystretch}{1.2}
\begin{tabular}{c|rrrrrrrr}
\hline $udsc\bar{c}$&
        $|1'>$  & $|1>$&$|2>$&$|3'>$&$|3>$&$|4>$&$|5>$\\\hline
$4273$&-0.54&0.06&-0.02&0.84&-0.05&-0.01&0.01\\
$4377$&-0.05&0.61&0.08&-0.12&-0.77&-0.15&-0.11\\
$4453$& 0.83&-0.03&0.10&0.52&-0.15&-0.09&0.03\\
$4469$&-0.07&-0.17&-0.20&-0.05&-0.11&-0.95&-0.09\\
$4494$&-0.02&0.46&0.64&-0.02&0.40&-0.30&0.36\\
$4576$&0.14&0.61&-0.55&0.06&0.45&-0.03&-0.31\\
$4649$&0.03&0.08&-0.48&-0.02&-0.11&0.02&0.87\\
\hline $uudc\bar{c}$& $|1>$&$|2>$&$|3>$&$|4>$&$|5>$\\\hline
$4267$& 0.61&0.11&-0.77&-0.03&-0.12\\
$4363$& 0.31&0.37&0.24&0.82&0.17\\
$4377$& 0.36&0.57&0.34&-0.56&0.34\\
$4471$& 0.63&-0.57&0.45&-0.05&-0.26\\
$4541$& 0.07&-0.44&-0.15&0.03&0.88\\
\hline
\end{tabular}
\end{table}

\begin{table}[h!]
\caption{The mixing coefficients of the states with spin-parity
$1/2^-$ under the $FS$ interaction. \label{coef}}
\renewcommand\tabcolsep{0.15cm}
\renewcommand{\arraystretch}{1.2}
\begin{tabular}{c|rrrrrrrr}
\hline $udsc\bar{c}$&
$|1'>$&$|1>$&$|2>$&$|3'>$&$|3>$&$|4>$&$|5>$\\\hline
$4084$&-0.03&-0.75&-0.66&0&0&0&0\\
$4154$& 0&0&0&0.39&-0.70&-0.58&0.12\\
$4160$& 0.95&-0.22&0.21&0&0&0&0\\
$4171$& 0&0&0&0.92&0.35&0.18&-0.06\\
$4253$& 0&0&0&-0.03&0.42&-0.35&0.84\\
$4263$&-0.29&-0.62&0.73&0&0&0&0\\
$4278$& 0&0&0&0.07&-0.46&0.71&0.53\\
\hline $uudc\bar{c}$& $|1>$&$|2>$&$|3>$&$|4>$&$|5>$\\\hline
$3933$& 0.76&0.65&0&0&0\\
$4013$& 0&0&-0.78&-0.60&0.17\\
$4119$& 0&0&0.52&-0.47&0.71\\
$4136$& 0.64&-0.76&0&0&0\\
$4156$& 0&0&0.35&-0.65&-0.68\\\hline
\end{tabular}
\end{table}

\begin{table}[h!]
\caption{The mixing coefficients of the states with spin-parity
$1/2^-$ under the $Inst.$ interaction.
\label{coef}}
\renewcommand\tabcolsep{0.15cm}
\renewcommand{\arraystretch}{1.2}
\begin{tabular}{c|rrrrrrrr}
\hline $udsc\bar{c}$&
$|1'>$&$|1>$&$|2>$&$|3'>$&$|3>$&$|4>$&$|5>$\\\hline
$4209$& 0.99&-0.07&0.08&0&0&0&0\\
$4216$& 0&0&0&0.97&0.12&0.02&0.19\\
$4277$&-0.04&-0.94&-0.35&0&0&0&0\\
$4295$& 0&0&0&-0.19&0.86&-0.21&0.42&\\
$4360$& -0.10&-0.34&0.93&0&0&0&0\\
$4362$&0&0&0&-0.07&0.13&0.97&0.19\\
$4416$&0&0&0&-0.10&-0.47&-0.12&0.87\\
\hline $uudc\bar{c}$& $|1>$&$|2>$&$|3>$&$|4>$&$|5>$\\\hline
$4089$& 0.94&0.35&0&0&0\\
$4096$& 0&0&0.86&-0.20&0.47\\
$4157$& 0&0&0.11&0.97&0.20\\
$4175$& -0.35&0.94&0&0&0\\
$4242$& 0&0&-0.50&-0.12&0.86\\
\hline
\end{tabular}
\end{table}

For the lowest spatial excited states, one quark should be in
$p$-wave, which results in a positive parity for the five quark
system. For the $udsc\bar c$ system, there are thirty four wave
functions with spin-parity $1/2^+$ and $3/2^+$,
twenty two with $5/2^+$ and four with $7/2^+$.
Similarly, there are too many states for $uudc \bar c$ system. Here,
ten of all states with spin-parity $1/2^+$, five lowest
states with spin-parity $3/2^+$, five lowest states with
$5/2^+$, and all the states with spin-parity
$7/2^+$ are listed Table~\ref{int3p} in terms of the energy.

\begin{table}[h!]
\caption{Energies (in unit of MeV)  of positive parity (L=1)
$qqqc\bar{c}$ states with quantum numbers of $N^*$- and
$\Lambda^*$-resonances under three kinds of interaction, with
configuration mixing considered.\label{int3p}}
\renewcommand\tabcolsep{0.25cm}
\renewcommand{\arraystretch}{1.}
\begin{tabular}{c|cc|cc|cc}
    \hline  &\multicolumn{2}{c|}{CM} & \multicolumn{2}{c|}{FS}
    & \multicolumn{2}{c}{$Inst.$}
\\
\hline$J^P$ &$udsc\bar{c}$ & $uudc\bar{c}$ & $udsc\bar{c}$
&$uudc\bar{c}$ & $udsc\bar{c}$&$uudc\bar{c}$ \\\hline
${\frac{1}{2}}^{+}$&$ 4622$&$ 4456$  &$4291$  &$4138$   & $4487$  &$4396$   \\
${\frac{1}{2}}^{+}$&$ 4636$&$ 4480$  &$4297$  &$4140$   & $4501$  &$4426$   \\
${\frac{1}{2}}^{+}$&$ 4645$&$ 4557$  &$4363$  &$4238$   & $4520$  &$4426$   \\
${\frac{1}{2}}^{+}$&$ 4658$&$ 4581$  &$4439$  &$4320$   & $4540$  &$4470$   \\
${\frac{1}{2}}^{+}$&$ 4690$&$ 4593$  &$4439$  &$4367$   & $4557$  &$4482$   \\
${\frac{1}{2}}^{+}$&$ 4696$&$ 4632$  &$4467$  &$4377$   & $4587$  &$4490$   \\
${\frac{1}{2}}^{+}$&$ 4714$&$ 4654$  &$4469$  &$4404$   & $4590$  &$4517$   \\
${\frac{1}{2}}^{+}$&$ 4728$&$ 4676$  &$4486$  &$4489$   & $4614$  &$4518$   \\
${\frac{1}{2}}^{+}$&$ 4737$&$ 4714$  &$4492$  &$4508$   & $4616$  &$4549$   \\
${\frac{1}{2}}^{+}$&$ 4766$&$ 4720$  &$4510$  &$4515$   & $4626$  &$4566$   \\
\hline
${\frac{3}{2}}^{+}$&$ 4623$&$ 4457$  &$4291$  &$4138$   & $4487$  &$4396$   \\
${\frac{3}{2}}^{+}$&$ 4638$&$ 4515$  &$4297$  &$4140$   & $4501$  &$4426$   \\
${\frac{3}{2}}^{+}$&$ 4680$&$ 4561$  &$4363$  &$4238$   & $4520$  &$4426$   \\
${\frac{3}{2}}^{+}$&$ 4692$&$ 4582$  &$4439$  &$4320$   & $4540$  &$4470$   \\
${\frac{3}{2}}^{+}$&$ 4695$&$ 4625$  &$4439$  &$4367$   & $4557$  &$4482$   \\
\hline
${\frac{5}{2}}^{+}$&$ 4705$&$ 4539$  &$4297$  &$4140$   & $4501$  &$4426$   \\
${\frac{5}{2}}^{+}$&$ 4719$&$ 4649$  &$4439$  &$4320$   & $4540$  &$4470$   \\
${\frac{5}{2}}^{+}$&$ 4773$&$ 4689$  &$4467$  &$4367$   & $4587$  &$4482$   \\
${\frac{5}{2}}^{+}$&$ 4793$&$ 4696$  &$4486$  &$4404$   & $4615$  &$4490$   \\
${\frac{5}{2}}^{+}$&$ 4821$&$ 4710$  &$4492$  &$4515$   & $4632$  &$4517$   \\
\hline
${\frac{7}{2}}^{+}$&$ 4945$&$ 4841$  &$4638$  &$4508$   & $4698$  &$4566$   \\
${\frac{7}{2}}^{+}$&$ 4955$&$ 4862$  &$4671$  &$4551$   & $4712$  &$4634$   \\
${\frac{7}{2}}^{+}$&$ 4974$&$ 4919$  &$4705$  &$4587$   & $4765$  &$4669$   \\
${\frac{7}{2}}^{+}$&$ 5010$&         &$4759$  &         & $4797$  &         \\
\hline
\end{tabular}
\end{table}

While in the negative parity sector there are three subspaces for
$1/2^-$, $3/2^-$ and $5/2^-$,
respectively, for the positive parity sector, there are four
subspaces for $1/2^+$, $3/2^+$,
$5/2^+$ and $7/2^+$, respectively. In the
process of the calculation, we take the $L$-$S$ coupling scheme with
standard Clebsch-Gordan coefficients of the angular momentum~\cite{pdg}. For the
flavor-spin and instanton-induced interactions, due to the ignoring
of quark-antiquark interaction, the $1/2^+$ and
$3/2^+$ states of the same configuration
$[f]_{FS}[f]_F[f]_S$ degenerate.  In the $CM$ model, the two states of
the same configuration but different four quark angular momentum
$\tilde{J}$ have a small splitting magnitude of several MeV as shown
in Table~\ref{int3p}. Here only the masses of several lower energy
states, which are more interesting to us, are listed in
Table~\ref{int3p}.

The non-zero off-diagonal matrix elements introduce the mixture of the
configurations with the same quantum number. The different hyperfine
interactions give different admixture of configurations of certain
state, which will result in different patterns of the
electromagnetic and strong decays. The mixing effect has been
explored in light quark sector, such as the decay of nucleon
resonances $N^*(1440)$~\cite{li} and  $N^*(1535)$~\cite{an2}.

For the $udsc\bar{c}$ system, in the $CM$ model without $q\bar{q}$
interaction, the SU(3) flavor singlet with hidden charm, which has
four quark configuration ${[211]_F}^{'}[31]_{CS}[211]_C[22]_S$, is
dominant in the lowest energy state, with a small admixture of
$[211]_F[31]_{CS}[211]_C[22]_S$. The mixing of the two
configurations is due to the flavor dependence of the $C_{i,j}$.
After considering the $q\bar{q}$ interaction in $CM$ model, the
configuration ${[211]_F}^{'}[31]_{CS}[211]_C[31]_S$ $(\sim72\%)$
becomes the dominant wave function component, with a strong
admixture of ${[211]_F}^{'}[31]_{CS}[211]_C[22]_S$ $(\sim27\%)$, as
shown in Table~\ref{coec}. The $q\bar{q}$ interaction leads to a
further mixing of the two spin symmetry configurations of $[22]_S$
and $[31]_S$, besides the flavor symmetry breaking effects. In the
FS model, the lowest state has a dominant four-quark configuration
$[31]_{FS}[211]_F[22]_S$ $(\sim42\%)$, with a strong admixtures of
$[31]_{FS}[31]_F[22]_S$ and  $[31]_{FS}[211]^{'}_F[22]_S$, as shown
in Table~\ref{coef}. In the $Inst$ model, the lowest state
predominantly has the configuration
$[211]_C[31]_{FS}{[211]_F}^{'}[22]_S$, which is the same as the $CM$
case without $q\bar{q}$ interaction.

For the $uudc\bar{c}$ system, there is no hidden charm SU(3) flavor
singlet state. In the $CM$ model after taking into account the
$q\bar{q}$ interaction, the lowest energy state is mainly the
admixture of $[211]_F[31]_{CS}[211]_C[31]_S$ $(\sim67\%)$ and
$[211]_F[31]_{CS}[211]_C[22]_S$ $(\sim27\%)$, as shown in
Table~\ref{coec}. In the $FS$ model, the lowest state is the
four-quark configuration $[31]_{FS}[211]_F[22]_S$ $(\sim52\%)$, with
a strong admixture of $[31]_{FS}[31]_F[22]_S$ $(\sim42\%)$. In the
present $Inst.$ model, assuming phenomenologically that the $'$t
Hooft's force also operates between a light and a charm quark, the
configuration $[211]_F[31]_{CS}[211]_C[22]_S$ should be the lowest,
as the spin $[22]_S$ and flavor $[211]_F$ contain more
antisymmetrized quark pairs. In the $Inst$ model, if it is assumed
that the light quark and charm quark decouples, the
$[211]_F[31]_{CS}[211]_C[22]_S$ and $[31]_F[31]_{CS}[211]_C[22]_S$
states degenerate and should be the lowest.

If the the flavor $SU(3)$ symmetry is restored and the light quark
and charm quark decouples, the $udsc\bar{c}$ is lower than the
$uudc\bar{c}$. For the positive parity $udsc\bar{c}$ states, under
the CM interaction with the $q\bar q$ interaction, the lowest state
has predominantly the four-quark configuration
$[31]_F[31]_{CS}[211]_C[31]_S$, with a strong admixture of the
configurations $[22]_F[31]_{CS}[211]_C[22]_S$ and
$[31]_F[31]_{CS}[211]_C[22]_S$. In the FS model, the lowest positive
parity state has predominantly the configuration
$[4]_{FS}[22]_C[22]_S$. The $Inst$ model predicts that the lowest
state is the configuration $[1111]_F[31]_{CS}[211]_C[22]_S$, which
can form the SU(3) flavor singlet state when combined with the
antiquark.

Different hyperfine interactions predict different configurations
for the lowest five quark states, which will result in different
decay patterns and can be checked by future experiments.

\section{Summary and discussions}
\label{sec:end}

In this work we have estimated the low-lying energy levels of the
five quark systems $uudc\bar{c}$ and $udsc\bar{c}$ with the hidden
charm by using the three kinds of hyperfine interactions. The hidden
charm states are obtained by diagonalizing the hyperfine
interactions in each subspace with the same spin-parity. For the
colormagnetic interaction, flavor-spin-dependent interaction and
$Inst.$-induced interaction, all the models predict that the lowest
states of the five quark systems $udsc\bar{c}$  and $uudc\bar{c}$
have the spin-parity $1/2^-$. The absolute value of the
negative hyperfine energy for the configuration
$[4]_{FS}[22]_F[22]_S$ in the positive parity sector is larger than
the case of the $[31]_{FS}{[211]_F}^{'}[22]_S$ in the negative
parity sector. But this difference cannot overcome the orbital
excited energy of the $P$-wave five quark system. This is in contrast
with the situation in the light flavor sector with the chiral
hyperfine interaction~\cite{Helminen}, due to the fact that the
hyperfine splitting depends on the quark masses and gets weak for
heavy quarks. In addition, for the flavor-spin interaction, the
lowest $uudc\bar{c}$ state has negative parity, which is opposite to
the lowest positive parity state of $uudd\bar{c}$ system containing
only one heavy antiquark~\cite{cbar}. The four quarks $uudd$ with
colored quark cluster configuration $[31]_X[4]_{FS}[22]_F[22]_S$ are
strongly attractive due to the diquark structure $[ud][ud]\bar c$.
However, the $c$ quark in the diquarks $[ud][uc]$ with the same
flavor-spin symmetry reduces to a large extent the hyperfine
interaction energy. The instanton-induced interaction only operates
on the color sextet and antitriplet diquark, and thus favors as well
the similar diquark structure. The $P$-wave diquark-triquark structure
$[ud][ud\bar c]$ is discussed under the colormagnetic
interaction~\cite{kl} and is almost as low as the $[ud][ud]\bar
c$~\cite{kim}. It would be of interests to study the configurations
of $[ud][uc\bar c]$ and $[ud][sc\bar c]$.

The coupled-channel unitary approach~\cite{wu} predicted that the
bound state $\bar{D}_s\Lambda_c$ is $30\sim50$ MeV lower than the
bound state $\bar{D}\Sigma_c$. In the chiral quark model
\cite{wangwl}, there only exists the bound state $\bar{D}\Sigma_c$.
In the present model, for the colored-cluster picture with three
kinds of the residual interactions, the lowest $udsc\bar{c}$ system
is heavier than the $uudc\bar{c}$ system. So the meson-baryon
picture and the penta-quark picture give different prediction on the
mass order of the super-heavy $N^*$ and $\Lambda^*$ with hidden
charm.

In the $CM$ model, the lowest $1/2^-$ and $3/2^-$
states, corresponding to the same four-quark configuration, are
split by the quark-antiquark interaction. And the $3/2^-$
state of the $udsc\bar{c}$ and $uudc\bar{c}$ system is about
$150~MeV$ heavier than the corresponding $1/2^-$ state. In
the $FS$ and $Inst.$ models, due to the lack of the quark-antiquark
interaction, the two states degenerate.

In addition, we have also discussed the admixture pattern of the
configurations with the same quantum numbers. The quark mass
difference and quark-anti-quark interaction are the two sources of
generating the configuration mixing, and the latter more important
for the configuration mixing and mass splitting of penta-quark
states. Since various configurations will result in different
electromagnetic and the strong decays, the study of the decay
properties may provide a good test of the models.

Experimental observation of the super-heavy $N^*$ and $\Lambda^*$
with hidden charm and their decay properties from $p\bar p$ reaction
at PANDA and $ep$ reaction at JLab 12 GeV upgrade are of great
interests for our understanding dynamics of strong interaction.

\section*{Acknowledgements}
This work is supported by the National Natural Science Foundation
of China (Grant Nos. 10875133, 10821063, 10905077, 10925526, 11035006 and 11147197), the Ministry of Education
of China (the project sponsored by SRF for ROCS, SEM under Grant No.
HGJO90402) and Chinese Academy of Sciences (the Special Foundation
of President under Grant No. YZ080425).

\begin{appendix}


\section{The wave functions for four quark subsystem}\label{wf4q}

\subsection{Flavor and spin couplings}

Take the decomposition of the flavor-spin configuration
$[31]_{FS}[211]_F [22]_S$ as an example,
\begin{eqnarray}
|[31]_{FS1}\rangle&=&\frac{1}{\sqrt{2}}\{|[211]\rangle_{F1}|[22]\rangle_{S1}+
|[211]\rangle_{F2}|[22]\rangle_{S2}\}\ ,\nonumber\\
\end{eqnarray}
\begin{eqnarray}
|[31]_{FS2}\rangle&=&\frac{1}{2}\{-\sqrt{2}|[211]\rangle_{F3}|[22]\rangle_{S2}
+ |[211]\rangle_{F2}|[22]\rangle_{S2}\nonumber\\&-&
|[211]\rangle_{F1}|[22]\rangle_{S1}\}\
\end{eqnarray}
\begin{eqnarray}
|[31]_{FS3}\rangle&=&\frac{1}{2}\{[211]\rangle_{F1}|[22]\rangle_{S2}
+ |[211]\rangle_{F2}|[22]\rangle_{S1} \nonumber\\&+&
|[211]\rangle_{F3}|[22]\rangle_{S1}\}\
\end{eqnarray}

\subsection{The flavor wave function of four quark subsystem $uudc$}
\label{uudc}

The explicit forms of the flavor symmetry $[211]_{F}$
\begin{eqnarray}
|[211]_{F1}\rangle&=&\frac{1}{4}\{2uudc-2uucd-duuc-uduc\nonumber\\
&-&cudu-ucdu+cuud+ducu\,\nonumber\\
&+&ucud+udcu\}\,,
\end{eqnarray}
\begin{eqnarray}
|[211]_{F2}\rangle&=&\frac{1}{\sqrt{48}}\{3uduc
-3duuc+3cuud\nonumber\\
&-&3ucud
+2dcuu-2cduu-cudu\,\nonumber\\
&+&ucdu+ducu-udcu\}\,,
\end{eqnarray}
\begin{eqnarray}
|[211]_{F3}\rangle&=&\frac{1}{\sqrt{6}}\{cudu+udcu
+dcuu\nonumber\\
&-&ucdu-ducu-cduu\}\,,
\end{eqnarray}

The explicit forms of the flavor symmetry $[22]_{F}$
\begin{eqnarray}
|[22]_{F_{1}}\rangle&=&
\frac{1}{\sqrt{24}}[2uudc+2uucd+2dcuu+2cduu\nonumber\\
&-&duuc-uduc-cudu- ucdu-cuud\nonumber\\
&-&ducu-ucud-udcu]\, , \label{combf1}
\end{eqnarray}
\begin{eqnarray}
|[22]_{F_{2}}\rangle&=&\!
\frac{1}{\sqrt{8}}[uduc+cudu+ducu+ucud\nonumber\\
&-&duuc-ucdu-cuud-udcu], \quad
\end{eqnarray}

The explicit forms of the flavor symmetry $[31]_{F}$
\begin{eqnarray}
|[31]_{F_{1}}\rangle \!&=& \frac{1}{\sqrt{18}}[2uucd+2cuud
+2ucud-cudu-ucdu\nonumber\\
&-&ducu-udcu-dcuu-cduu]\, , \label{combf4}
\end{eqnarray}
\begin{eqnarray}
|[31]_{F_{2}}\rangle \!&=& \frac{1}{12}[6uudc-3duuc
-3uduc-4dcuu-4cduu\nonumber\\
&+&5cudu+5ucdu+2uucd-cuud-ducu\nonumber\\
&-&ucud-udcu]\, , \label{combf7}
\end{eqnarray}
\begin{eqnarray}
|[31]_{F_{3}}\rangle \!&=& \frac{1}{\sqrt{48}} [-3duuc +3uduc-3ducu
+3udcu\nonumber\\
&-&2dcuu+2cduu-cudu+ucdu -cuud\nonumber\\
&+&ucud]\ . \label{combf8}
\end{eqnarray}

\subsection{The flavor wave function of four quark subsystem $udsc$}
\label{uusc}

The explicit forms of the flavor symmetry $[31]_{F}$:
\begin{eqnarray}
|[31]_{F_{1}}\rangle  &=&
\frac{1}{\sqrt{12}}[ucsd-cdsu+uscd-dcsu+sucd\nonumber\\
&-&sdcu+cusd-dscu+scud-scdu\nonumber\\
&+&csud-csdu]\, , \label{combf5}
\end{eqnarray}
\begin{eqnarray}
|[31]_{F_{2}}\rangle
&=&\frac{1}{\sqrt{96}}\{3(usdc-sduc+ucds-cdus\nonumber\\
&+&sudc-dsuc+cuds-dcus)\nonumber\\
&+&2(scdu-scud+csdu-csud)\nonumber\\
&+&ucsd-cdsu+uscd-dcsu+sucd\nonumber\\
&-&sdcu+cusd-dscu\}\, ,
\end{eqnarray}
\begin{eqnarray}
|[31]_{F_{3}}\rangle &=&
\frac{1}{\sqrt{32}}\{2(udsc-dusc+udcs-ducs)\nonumber\\
&+&sduc+usdc+cdsu+ucsd+cdus\nonumber\\
&+&ucds+uscd+sdcu-dcsu-sucd-sudc\,\nonumber\\
&-&cuds-cusd-dsuc-dscu-dcus\}\,\quad \label{combfl}
\end{eqnarray}

The explicit forms of the flavor symmetry $[211]_{F}$:
\begin{eqnarray}
|[211]_{F1}\rangle&=&\frac{1}{4\sqrt{6}}\{3(sudc-sduc+usdc-dsuc\nonumber\\
&+&sdcu-sucd+dscu-uscd)\nonumber\\
&+&2(csud-csdu+scud-scdu)\,\nonumber\\
&+&cusd-cdsu+cdus-cuds\nonumber\\
&+&dcus-ucds+ucsd-dcsu\}\, ,
\end{eqnarray}
\begin{eqnarray}
|[211]_{F2}\rangle&=& \frac{1}{12\sqrt{2}}\{6(udsc
-dusc)\nonumber\\&+&5(dcsu-cdsu
+cusd-ucsd )\nonumber\\
&+&4(scdu-csdu+csud-scud)\nonumber\\
&+&3(sduc-dsuc+usdc-sudc)\nonumber\\
&+&2(ducs-udcs)+cuds-ucds+sucd\nonumber\\
&-&sdcu+dscu-cdus+dcus-uscd\}\, ,
\end{eqnarray}
\begin{eqnarray}
|[211]_{F3}\rangle&=&
\frac{1}{6}\{2(udcs+cuds+dcus-ducs\nonumber\\
&-&cdus-ucds)+dcsu+uscd+scud\nonumber\\
&+&sdcu+cusd+csdu-cdsu\,\nonumber\\
&-&ucsd-sucd-scdu-csud-dscu\}
\end{eqnarray}

The explicit forms of the flavor symmetry $[211]^{'}_{F}$

\begin{eqnarray}
|[211]^{'}_{F1}\rangle
&=&\frac{1}{2\sqrt{3}}\{cdsu-ucsd+ucds-cdus\nonumber\\
&+&cuds-scdu+dcsu-cusd\nonumber\\&+&dscu-csdu+csud-dcus\}\nonumber\\
\end{eqnarray}
\begin{eqnarray}
|[211]^{'}_{F2}\rangle
&=&\frac{1}{6}\{2(udcs-sdcu+dscu-ducs+sucd-uscd)\nonumber\\&+&cdus-cdsu+ucds-ucsd
+scud\nonumber\\&-&scdu+dcsu-cuds+cusd\nonumber\\&-&csud+csdu-dcus\}\nonumber\\
\end{eqnarray}
\begin{eqnarray}
|[211]^{'}_{F3}\rangle
&=&\frac{1}{6\sqrt{2}}\{3(udsc+sudc+dsuc-dusc\nonumber\\&-&sduc-usdc)+cdsu+
ucsd+udcs+sucd\nonumber\\&+&scdu+cuds+csud+dscu+dcus-ducs\nonumber\\
&-&cdus-ucds-dcsu-uscd\nonumber\\&-&scud-sdcu-cusd-csdu\}\nonumber\\
\end{eqnarray}

\subsection{The wave function of spin symmetry of four quark subsystem}

The wave functions for spin symmetry $[22]_{S}$,
\begin{eqnarray}
|[22]\rangle_{S1}&=&\frac{1}{\sqrt{12}}\{2
|\uparrow\uparrow\downarrow\downarrow\rangle
+2|\downarrow\downarrow\uparrow\uparrow\rangle
-|\downarrow\uparrow\uparrow\downarrow\rangle
-|\uparrow\downarrow\uparrow\downarrow\rangle
\nonumber\\
&-&|\downarrow\uparrow\downarrow\uparrow\rangle
-|\uparrow\downarrow\downarrow\uparrow\rangle\}\,,\\
|[22]\rangle_{S2}&=&\frac{1}{2}\{
|\uparrow\downarrow\uparrow\downarrow\rangle
+|\downarrow\uparrow\downarrow\uparrow\rangle
-|\downarrow\uparrow\uparrow\downarrow\rangle
-|\uparrow\downarrow\downarrow\uparrow\rangle\}\,.
\end{eqnarray}
More can be found in Ref.~\cite{an1}.

\end{appendix}

\end{document}